\documentclass[aps,prd,onecolumn,amsmath,amssymb,nofootinbib,superscriptaddress,10pt,floatfix]{revtex4-2}
\usepackage{graphicx}
\usepackage{amsmath,amssymb,amsfonts,mathrsfs}
\usepackage{bm}
\usepackage{xcolor}
\usepackage{colortbl}
\usepackage{booktabs}
\usepackage{array}
\usepackage{subcaption}
\usepackage{hyperref}
\usepackage{placeins}
\usepackage{tabularx}
\usepackage{orcidlink}     % needed for \orcidlink badges in author block
\hypersetup{colorlinks=true,linkcolor=blue,citecolor=blue,urlcolor=blue}

\newcolumntype{C}{>{\centering\arraybackslash}X}
\newcolumntype{L}{>{\raggedright\arraybackslash}X}
\newcolumntype{R}{>{\raggedleft\arraybackslash}X}

\numberwithin{equation}{section}

\newcommand{\func}[1]{\operatorname{#1}}

\begin{document}

\title{Charged Axially Symmetric Exponential Metric: Exact Solutions to the
Einstein-Maxwell Equations}
\author{S. Habib Mazharimousavi\,\orcidlink{0000-0002-7035-6155}}
\email{habib.mazhari@emu.edu.tr}
\affiliation{Department of Physics, Faculty of Arts and Sciences, Eastern Mediterranean
University, Famagusta, North Cyprus via Mersin 10, T\"{u}rkiye}
\date{\today }

\begin{abstract}
We construct and analyze charged and magnetized extensions of the axially
symmetric exponential metric, that is known as the Curzon-Chazy spacetime
(CCS) within the Weyl class. Working within the Einstein-Maxwell framework,
we first derive an exact charged axially symmetric exponential metric by
directly solving the coupled field equations for a dyonic monopole
configuration. The solution is shown to reduce smoothly to the vacuum CCS in
the neutral limit and to the extremal Reissner-Nordstr\"{o}m spacetime in a
special parameter regime. We then rederive the same charged geometry using
the Harrison transformation in the Ernst formalism, establishing the
equivalence between the direct and solution-generating approaches.
Subsequently, we apply the magnetic Harrison transformation to obtain a
Melvin-type magnetized deformation of the spacetime. Finally, by
implementing the Ehlers transformation, we generate a stationary but
nonrotating swirling geometry endowed with a nonvanishing NUT charge, which
we identify invariantly via the Komar dual mass. A further magnetic Harrison
transformation yields a magnetized swirling (NUT-type) spacetime. These
results demonstrate how electric, magnetic, and gravitomagnetic charges
arise systematically from the CCS through exact solution-generating
techniques.
\end{abstract}

\maketitle

\section{Introduction}

In a recent study \cite{1}, we investigated the axially symmetric
generalization of the exponential metric \cite{2,3,4,5,6,7} and the
Curzon-Chazy spacetime (CCS) \cite{Cur,Cha}. The latter is a static vacuum
solution of Einstein's equations that modifies the Schwarzschild metric by
incorporating an exponential dependence in the metric function. The line
element of the one-parameter static axially symmetric exponential metric
i.e., CCS is given by 
\begin{equation}
ds^{2}=-e^{-\frac{2M}{r}}dt^{2}+e^{\frac{2M}{r}}\left\{ e^{-\frac{M^{2}\sin
^{2}\theta }{r^{2}}}\left( dr^{2}+r^{2}d\theta ^{2}\right) +r^{2}\sin
^{2}\theta \,d\phi ^{2}\right\} ,  \label{I1}
\end{equation}%
in which the only parameter i.e., $M$ is the ADM of the solution. This
metric is an exact singular solution of the vacuum Einstein equations and
represents an axially symmetric extension of the spherically symmetric
exponential metric, commonly known as the Yilmaz exponential metric (YEM),
whose line element is 
\begin{equation}
ds^{2}=-e^{-\frac{2M}{r}}dt^{2}+e^{\frac{2M}{r}}\left( dr^{2}+r^{2}d\theta
^{2}+r^{2}\sin ^{2}\theta \,d\phi ^{2}\right) .  \label{I2}
\end{equation}%
The YEM is supported by an exotic energy-momentum tensor of the form 
\begin{equation}
T_{\mu }^{\nu }=\frac{M^{2}e^{-\frac{2M}{r}}}{r^{4}}\func{diag}\left(
1,-1,1,1\right) .  \label{I3}
\end{equation}%
The YEM has a long history, which is briefly reviewed in \cite{1} (see also
related works \cite{2,3,4,5,6,7}). More recently, Boonserm \emph{et al.} 
\cite{8} demonstrated that, interestingly, the YEM describes an asymmetric
traversable wormhole with its throat located at $r=M$ (for traversable
wormholes, see \cite{9,10,11,12,13}). In contrast, the CCS is a vacuum
solution that, in the asymptotic region, mimics the Schwarzschild spacetime
expressed in isotropic coordinates. Specifically, as $r\rightarrow \infty $,
Eq.~(\ref{I1}) approaches 
\begin{equation}
ds^{2}\rightarrow -\left( 1-\frac{2M}{r}\right) dt^{2}+\left( 1+\frac{2M}{r}%
\right) \left( dr^{2}+r^{2}d\theta ^{2}+r^{2}\sin ^{2}\theta \,d\phi
^{2}\right) .  \label{I4}
\end{equation}%
Having established the properties of the CCS, we are now interested in
constructing its charged counterpart. In this paper, we initiate our study
by formulating the field equations within the Einstein-Maxwell (EM)
framework and solving them to obtain a charged generalization of the CCS.

On the other hand, solution-generating techniques have played a central role
in the study of exact solutions in general relativity since the mid-20th
century. A major breakthrough was the discovery of the Ernst formulation of
the stationary, axisymmetric Einstein and EM equations, which recasts the
tensorial field equations into a pair of complex scalar equations governing
the Ernst potentials $\mathcal{E}$ and $\Phi $ \cite{Ernst1,Ernst2}. This
framework revealed hidden symmetry structures of the field equations, laying
the groundwork for systematic generation of new electrovac solutions from
simpler \textit{seed}\ spacetimes. Among the earliest and most influential
of these methods is the Harrison transformation, introduced by B.~Kent
Harrison in 1968, which acts as a nonlinear mapping on Ernst potentials to
inject a given seed with electromagnetic charge \cite{Harrison1968} and has
since been utilized to construct magnetized black hole spacetimes such as
the Ernst-Melvin solutions \cite{ErnstWild1976,Ernst1976}. Complementing
this, J.~Ehlers explored a class of solution-generating transformations in
the same era that effectively mix the gravitational and twist sectors of the
Ernst potentials, introducing additional parameters that can be associated
with gravitomagnetic, or NUT, charges \cite{Ehlers1957}. These
transformations are part of a larger symmetry group of the stationary
axisymmetric EM field equations, including later formulations by Kinnersley
and Geroch and their extensions to incorporate electromagnetism \cite%
{Kinnersley1973,Geroch1971}. The combined action of Harrison and Ehlers
transformations (and their generalizations) has also been explored within
dilaton-axion and other extended gravity theories, underscoring the unifying
power of these methods in exact solution generation \cite%
{Galtsov1994,HerreraAguilar1998}.

Recent work continues to exploit and generalize these techniques in novel
directions. For example, a systematic classification of stationary
axisymmetric spacetimes generated via compositions of Harrison and Ehlers
transformations highlights their ability to produce new geometries combining
electromagnetic, gravitomagnetic, and external field effects \cite%
{Mixing2024}. Generalized Harrison transformations have been constructed in
modified EM models, preserving specific charge sectors while producing new
families of solutions such as black holes embedded in nonlinear
electrodynamic backgrounds \cite{Barrientos2025}. Other efforts focus on
refinements of the Ehlers transformation itself, such as enhanced maps that
act as gravitational dualities rotating mass and NUT charges in electrovac
settings \cite{EnhancedEhlers2020}. These developments reflect the enduring
importance of solution-generating symmetries in uncovering and understanding
the rich landscape of exact spacetimes in general relativity and its
extensions. In this work, we place Harrison and Ehlers transformations at
the heart of our construction of charged, magnetized, and swirling
extensions of the axially symmetric exponential metric.

\section{Charged CCS: A field equation approach}

We begin with the standard action for EM theory in natural units ($c=G=1$), 
\begin{equation}
S=\int d^{4}x\sqrt{-g}\left( R-\frac{1}{4}F_{\mu \nu }F^{\mu \nu }\right) ,
\label{1}
\end{equation}%
where $R$ is the Ricci scalar curvature, $g$ denotes the determinant of the
metric tensor $g_{\mu \nu }$, and $F_{\mu \nu }F^{\mu \nu }$ is the Maxwell
invariant, with $F_{\mu \nu }$ being the electromagnetic field strength
tensor. Variation of the action with respect to the metric $g_{\mu \nu }$
yields the Einstein field equations 
\begin{equation}
G_{\mu }^{\nu }=8\pi T_{\mu }^{\nu },  \label{2}
\end{equation}%
where $G_{\mu }^{\nu }$ is the Einstein tensor and $T_{\mu }^{\nu }$ is the
electromagnetic energy-momentum tensor, given by 
\begin{equation}
T_{\mu }^{\nu }=\frac{1}{4\pi }\left( F_{\mu \alpha }F^{\nu \alpha }-\frac{1%
}{4}F_{\alpha \beta }F^{\alpha \beta }\delta _{\mu }^{\nu }\right) .
\label{3}
\end{equation}%
Variation of the action with respect to the electromagnetic potential $%
A_{\mu }$ leads to the source-free Maxwell equations 
\begin{equation}
\nabla _{\mu }F^{\mu \nu }=0,  \label{4}
\end{equation}%
where $\nabla _{\mu }$ denotes the covariant derivative. In addition, the
Bianchi identity, which follows from the definition $F_{\mu \nu }=\partial
_{\mu }A_{\nu }-\partial _{\nu }A_{\mu }$, implies 
\begin{equation}
\nabla _{\lbrack \lambda }F_{\mu \nu ]}=0.  \label{5}
\end{equation}%
To explore solutions within this framework, we consider a line element
inspired by the exponential metric, 
\begin{equation}
ds^{2}=-\psi (r)\,dt^{2}+\frac{1}{\psi (r)}\left\{ \xi (r,\theta )\left(
dr^{2}+r^{2}d\theta ^{2}\right) +r^{2}\sin ^{2}\theta \,d\phi ^{2}\right\} ,
\label{6}
\end{equation}%
where $\psi (r)$ and $\xi (r,\theta )$ are functions to be determined from
the EM equations. For the electromagnetic field, we adopt the ansatz 
\begin{equation}
\mathbf{F}=E(r)\,dt\wedge dr+P\sin \theta \,d\theta \wedge d\phi ,  \label{7}
\end{equation}%
where $E(r)$ represents the radial electric field component and $P$ is a
constant corresponding to a magnetic monopole charge. This configuration
describes a combined electric and magnetic field generated by central
monopole sources.

The Hodge dual of the electromagnetic field tensor $\mathbf{F}$, denoted by $%
\star \mathbf{F}$, is given by 
\begin{equation}
\star \mathbf{F}=\frac{E(r)\,r^{2}}{\psi (r)}\sin \theta \,d\theta \wedge
d\phi -\frac{P\psi (r)}{r^{2}}\,dt\wedge dr.  \label{8}
\end{equation}%
It is straightforward to verify that $d\mathbf{F}=0$ (or $\nabla _{\lbrack
\lambda }F_{\mu \nu ]}=0$) and $d\star \mathbf{F}=0$ (or $\nabla _{\mu
}F^{\mu \nu }=0$) are satisfied, provided that $P$ is a constant
representing the magnetic monopole charge and that the electric field
function $E(r)$ is chosen as 
\begin{equation}
E(r)=\frac{Q\,\psi (r)}{r^{2}},  \label{9}
\end{equation}%
where $Q$ is an integration constant corresponding to the electric monopole
charge. The Maxwell invariant is then computed as 
\begin{equation}
F_{\alpha \beta }F^{\alpha \beta }=\frac{2\left( P^{2}-Q^{2}\right) \psi ^{2}%
}{r^{4}\xi },  \label{10}
\end{equation}%
and consequently, the energy-momentum tensor takes the form 
\begin{equation}
T_{\mu }^{\nu }=\frac{\left( P^{2}+Q^{2}\right) \psi ^{2}}{8\pi r^{4}\xi }%
\func{diag}\left( -1,-1,1,1\right) .  \label{11}
\end{equation}%
For the line element (\ref{6}), the nonvanishing components of the Einstein
tensor $G_{\mu }^{\nu }$ are found to be 
\begin{equation}
G_{t}^{t}=\frac{\frac{5}{4}\frac{\psi ^{\prime 2}}{\psi }-\psi ^{\prime
\prime }-\frac{2}{r}\psi ^{\prime }}{\xi }+\frac{\psi \left( \frac{\partial
^{2}\xi }{\partial \theta ^{2}}+r^{2}\frac{\partial ^{2}\xi }{\partial r^{2}}%
+r\frac{\partial \xi }{\partial r}\right) }{2\xi ^{2}r^{2}}-\frac{\psi \left[
\left( \frac{\partial \xi }{\partial \theta }\right) ^{2}+r^{2}\left( \frac{%
\partial \xi }{\partial r}\right) ^{2}\right] }{2\xi ^{3}r^{2}},  \label{12}
\end{equation}%
\begin{equation}
G_{r}^{r}=-\frac{1}{4}\frac{\psi ^{\prime 2}}{\psi \xi }+\frac{\psi \left(
r\sin \theta \frac{\partial \xi }{\partial r}-\cos \theta \frac{\partial \xi 
}{\partial \theta }\right) }{2r^{2}\xi ^{2}\sin \theta },  \label{13}
\end{equation}%
\begin{equation}
G_{\theta }^{\theta }=\frac{1}{4}\frac{\psi ^{\prime 2}}{\psi \xi }-\frac{%
\psi \left( r\sin \theta \frac{\partial \xi }{\partial r}-\cos \theta \frac{%
\partial \xi }{\partial \theta }\right) }{2r^{2}\xi ^{2}\sin \theta },
\label{14}
\end{equation}%
\begin{equation}
G_{\phi }^{\phi }=\frac{1}{4}\frac{\psi ^{\prime 2}}{\psi \xi }+\frac{\psi
\left( \frac{\partial ^{2}\xi }{\partial \theta ^{2}}+r^{2}\frac{\partial
^{2}\xi }{\partial r^{2}}+r\frac{\partial \xi }{\partial r}\right) }{2\xi
^{2}r^{2}}-\frac{\psi \left[ \left( \frac{\partial \xi }{\partial \theta }%
\right) ^{2}+r^{2}\left( \frac{\partial \xi }{\partial r}\right) ^{2}\right] 
}{2\xi ^{3}r^{2}},  \label{15}
\end{equation}%
and 
\begin{equation}
G_{\theta }^{r}=G_{r}^{\theta }=\frac{\psi \left( \sin \theta \frac{\partial
\xi }{\partial \theta }+r\cos \theta \frac{\partial \xi }{\partial r}\right) 
}{2r^{3}\xi ^{2}\sin \theta }.  \label{16}
\end{equation}%
Since the energy-momentum tensor is diagonal, the off-diagonal component $%
G_{\theta }^{r}$ must vanish, which leads to the condition 
\begin{equation}
\xi (r,\theta )=\xi \!\left( \frac{\sin \theta }{r}\right) ,  \label{17}
\end{equation}%
indicating that $\xi $ depends only on the variable $z=\sin \theta /r$.
Moreover, the relation $T_{t}^{t}+T_{\phi }^{\phi }=0$ implies $%
G_{t}^{t}+G_{\phi }^{\phi }=0$, yielding the differential equation 
\begin{equation}
\xi (z)\,\xi ^{\prime \prime }(z)-\left( \xi ^{\prime }(z)\right)
^{2}=-2C_{0}\,\xi ^{2}(z),  \label{18}
\end{equation}%
where $C_{0}$ is an integration constant. Solving this equation, we obtain 
\begin{equation}
\xi (r,\theta )=\exp \!\left[ -C_{0}\left( \frac{\sin \theta }{r}\right)
^{2}+C_{1}\frac{\sin \theta }{r}+C_{2}\right] ,  \label{19}
\end{equation}%
with $C_{1}$ and $C_{2}$ being integration constants. Substituting this
expression into the $rr$ and $\theta \theta $ components of the EM
equations, we find that $C_{1}=0$, while the function $\psi (r)$ satisfies
the nonlinear differential equation 
\begin{equation}
-\left( \psi ^{\prime }(r)\right) ^{2}r^{4}+4C_{0}\psi ^{2}(r)+4\left(
P^{2}+Q^{2}\right) \psi ^{3}(r)=0.  \label{20}
\end{equation}%
Furthermore, the $tt$ and $\phi \phi $ components yield 
\begin{equation}
-4r^{4}\psi \psi ^{\prime \prime }+5r^{4}\left( \psi ^{\prime }\right)
^{2}-8r^{3}\psi \psi ^{\prime }+4\left( P^{2}+Q^{2}\right) \psi
^{3}-4C_{0}\psi ^{2}=0.  \label{21}
\end{equation}%
It is convenient, and without loss of generality, to set $C_{2}=0$. The
exact solution for $\psi (r)$ is then found to be 
\begin{equation}
\psi (r)=\frac{C_{0}}{P^{2}+Q^{2}}\left[ \left( \frac{e^{-2\sqrt{C_{0}}/r}+K%
}{e^{-2\sqrt{C_{0}}/r}-K}\right) ^{2}-1\right] ,  \label{22}
\end{equation}%
where $K$ is an integration constant. Requiring that the solution reduces to
CCS in the limit $P,Q\rightarrow 0$, we impose $\psi (r)\rightarrow
e^{-2M/r} $ and $\xi (r,\theta )\rightarrow e^{-(M\sin \theta /r)^{2}}$.
This yields 
\begin{equation}
\psi (r)=\left[ \cosh \!\left( \frac{M}{r}\right) +\frac{\mu }{M}\sinh
\!\left( \frac{M}{r}\right) \right] ^{-2},  \label{23}
\end{equation}%
and 
\begin{equation}
\xi (r,\theta )=e^{-\frac{M^{2}\sin ^{2}\theta }{r^{2}}},  \label{24}
\end{equation}%
where 
\begin{equation}
M^{2}=\mu ^{2}-\left( P^{2}+Q^{2}\right) .  \label{25}
\end{equation}%
Note that the solutions (\ref{23}) and (\ref{24}) are valid under the
condition 
\begin{equation*}
M^{2}=\mu ^{2}-p^{2}>0,
\end{equation*}%
where, for brevity, we have introduced $p^{2}=P^{2}+Q^{2}$. Consequently,
the corresponding line element takes the form 
\begin{multline}
ds^{2}=-\frac{dt^{2}}{\left[ \cosh \!\left( \frac{M}{r}\right) +\frac{\mu }{M%
}\sinh \!\left( \frac{M}{r}\right) \right] ^{2}}+  \label{26} \\
\left[ \cosh \!\left( \frac{M}{r}\right) +\frac{\mu }{M}\sinh \!\left( \frac{%
M}{r}\right) \right] ^{2}\left\{ e^{-\frac{M^{2}\sin ^{2}\theta }{r^{2}}%
}\left( dr^{2}+r^{2}d\theta ^{2}\right) +r^{2}\sin ^{2}\theta \,d\phi
^{2}\right\} .
\end{multline}%
Although the Ricci scalar vanishes, as expected from the traceless nature of
the electromagnetic energy-momentum tensor, the Kretschmann scalar diverges
as $r\rightarrow 0$, indicating the presence of a curvature singularity at
the origin. In the limit $(P^{2}+Q^{2})\rightarrow 0$, this solution reduces
to the vacuum CCS. Moreover, in the asymptotic region $r\rightarrow \infty $%
, the metric approaches 
\begin{equation}
ds^{2}=-\left( 1-\frac{2M}{r}+\frac{3\mu ^{2}-M^{2}}{r^{2}}\right)
dt^{2}+\left( 1+\frac{2M}{r}+\frac{\mu ^{2}+M^{2}}{r^{2}}\right) \left\{
\left( 1-\frac{M^{2}\sin ^{2}\theta }{r^{2}}\right) \left(
dr^{2}+r^{2}d\theta ^{2}\right) +r^{2}\sin ^{2}\theta \,d\phi ^{2}\right\} .
\label{28}
\end{equation}%
For comparison, consider the Reissner-Nordstr\"{o}m (RN) metric in standard
Schwarzschild coordinates, 
\begin{equation}
d\tilde{s}^{2}=-\left( 1-\frac{2\tilde{m}}{\tilde{r}}+\frac{\tilde{p}^{2}}{%
\tilde{r}^{2}}\right) d\tilde{t}^{2}+\left( 1-\frac{2\tilde{m}}{\tilde{r}}+%
\frac{\tilde{p}^{2}}{\tilde{r}^{2}}\right) ^{-1}d\tilde{r}^{2}+\tilde{r}%
^{2}\left( d\theta ^{2}+\sin ^{2}\theta \,d\phi ^{2}\right) ,  \label{29}
\end{equation}%
where $\tilde{m}$ and $\tilde{p}$ denote the mass and charge of the RN
spacetime. To express this metric in isotropic coordinates, we apply the
transformation $\tilde{r}=U(r)\,r$, with 
\begin{equation}
U(r)\,dr=\frac{d\tilde{r}}{\sqrt{1-\frac{2\tilde{m}}{\tilde{r}}+\frac{\tilde{%
p}^{2}}{\tilde{r}^{2}}}},  \label{30}
\end{equation}%
which yields the line element 
\begin{equation}
d\tilde{s}^{2}=-V(r)\,d\tilde{t}^{2}+U^{2}(r)\left\{ dr^{2}+r^{2}\left(
d\theta ^{2}+\sin ^{2}\theta \,d\phi ^{2}\right) \right\} .  \label{31}
\end{equation}%
From Eq.~(\ref{31}), one obtains 
\begin{equation}
\tilde{r}=\frac{r}{2}\left[ \left( 1+\frac{\tilde{m}}{r}\right) ^{2}-\frac{%
\tilde{p}^{2}}{r^{2}}\right] ,  \label{32}
\end{equation}%
leading to 
\begin{equation}
U(r)=\frac{1}{2}\left[ \left( 1+\frac{\tilde{m}}{r}\right) ^{2}-\frac{\tilde{%
p}^{2}}{r^{2}}\right] ,  \label{33}
\end{equation}%
and 
\begin{equation}
V(r)=\left( \frac{1-\frac{\tilde{m}^{2}}{r^{2}}+\frac{\tilde{p}^{2}}{r^{2}}}{%
\left( 1+\frac{\tilde{m}}{r}\right) ^{2}-\frac{\tilde{p}^{2}}{r^{2}}}\right)
^{2}.  \label{34}
\end{equation}%
Finally, upon rescaling $d\tilde{s}=2ds$ and $d\tilde{t}=2dt$, we arrive at 
\begin{equation}
ds^{2}=-\left( \frac{1-\frac{\tilde{m}^{2}}{r^{2}}+\frac{\tilde{p}^{2}}{r^{2}%
}}{\left( 1+\frac{\tilde{m}}{r}\right) ^{2}-\frac{\tilde{p}^{2}}{r^{2}}}%
\right) ^{2}dt^{2}+\left[ \left( 1+\frac{\tilde{m}}{r}\right) ^{2}-\frac{%
\tilde{p}^{2}}{r^{2}}\right] ^{2}\left\{ dr^{2}+r^{2}d\theta ^{2}+r^{2}\sin
^{2}\theta \,d\phi ^{2}\right\} ,  \label{35}
\end{equation}%
which is the RN metric in isotropic coordinates. Its asymptotic expansion
takes the form 
\begin{equation}
ds^{2}=-\left( 1-\frac{4\tilde{m}}{r}+\frac{8\tilde{m}^{2}+4\tilde{p}^{2}}{%
r^{2}}\right) dt^{2}+\left( 1+\frac{4\tilde{m}}{r}+\frac{6\tilde{m}^{2}-2%
\tilde{p}^{2}}{r^{2}}\right) \left\{ dr^{2}+r^{2}d\theta ^{2}+r^{2}\sin
^{2}\theta \,d\phi ^{2}\right\} .  \label{36}
\end{equation}%
Comparing this expression with the asymptotic form of our metric in Eq.~(\ref%
{28}), we observe that under the rescalings $2\tilde{m}=M$ and $2\tilde{p}=p$%
, the $g_{tt}$ components exhibit strong agreement. This indicates that,
asymptotically, the spacetime resembles a deformed isotropic RN geometry.

Furthermore, although the solution (\ref{22}) satisfies the field equations,
the constants $K$ and $C_{0}$ must be fixed appropriately to ensure an
asymptotic match with the Schwarzschild solution. In addition, the condition 
$\mu ^{2}-p^{2}\geq 0$ must be imposed. In the limiting case $M^{2}=\mu
^{2}-p^{2}\rightarrow 0$, the metric reduces to 
\begin{equation}
ds^{2}=-\left( 1+\frac{\mu }{r}\right) ^{-2}dt^{2}+\left( 1+\frac{\mu }{r}%
\right) ^{2}\left\{ dr^{2}+r^{2}d\theta ^{2}+r^{2}\sin ^{2}\theta \,d\phi
^{2}\right\} .  \label{37}
\end{equation}%
Finally, introducing the coordinate transformation $r=\tilde{r}-\mu $, the
metric reduces to the standard form of the extremal RN black hole, 
\begin{equation}
ds^{2}=-\left( 1-\frac{\mu }{\tilde{r}}\right) ^{2}dt^{2}+\left( 1-\frac{\mu 
}{\tilde{r}}\right) ^{-2}d\tilde{r}^{2}+\tilde{r}^{2}\left( d\theta
^{2}+\sin ^{2}\theta \,d\phi ^{2}\right) .  \label{38}
\end{equation}

\section{Harrison transformation}

Within the Ernst formalism \cite{Ernst1,Ernst2}, the EM equations for
stationary, axially symmetric spacetimes reduce to the coupled system 
\begin{equation}
\left( \Re \mathcal{E}+|\Phi |^{2}\right) \nabla ^{2}\mathcal{E}=\nabla 
\mathcal{E}\cdot \left( \nabla \mathcal{E}+2\Phi ^{\star }\nabla \Phi
\right) ,  \label{39}
\end{equation}%
and%
\begin{equation}
\left( \Re \mathcal{E}+|\Phi |^{2}\right) \nabla ^{2}\Phi =\nabla \Phi \cdot
\left( \nabla \mathcal{E}+2\Phi ^{\star }\nabla \Phi \right) ,  \label{40}
\end{equation}%
where the complex Ernst potentials are defined by 
\begin{equation}
\mathcal{E}=f-|\Phi |^{2}+i\chi ,  \label{41}
\end{equation}%
and the stationary axisymmetric line element takes the Weyl-Papapetrou form 
\begin{equation}
ds^{2}=-f(\rho ,z)\left( dt-\omega (\rho ,z)d\varphi \right) ^{2}+\frac{1}{%
f(\rho ,z)}\left[ e^{2\gamma (\rho ,z)}(d\rho ^{2}+dz^{2})+\rho ^{2}d\varphi
^{2}\right] .  \label{42}
\end{equation}%
The complex electromagnetic potential is given by 
\begin{equation}
\Phi (\rho ,z)=A_{t}(\rho ,z)+i\tilde{A}_{\varphi }(\rho ,z),  \label{43}
\end{equation}%
where the auxiliary magnetic potential $\tilde{A}_{\varphi }(\rho ,z)$
satisfies 
\begin{equation}
\hat{\varphi}\times \nabla \tilde{A}_{\varphi }=\rho ^{-1}f\left( \nabla
A_{\varphi }+\omega \nabla A_{t}\right) .  \label{44}
\end{equation}%
Also, the twist potential $\chi (\rho ,z)$ is defined through 
\begin{equation}
\hat{\varphi}\times \nabla \chi =-\rho ^{-1}f^{2}\nabla \omega -2\hat{\varphi%
}\times \Im \!\left( \Phi ^{\star }\nabla \Phi \right) .  \label{45}
\end{equation}%
Herein the operator Del is defined in 3-dimensional Euclidean cylindrical
coordinates i.e., $\nabla =\hat{\rho}\frac{\partial }{\partial \rho }+\hat{%
\varphi}\frac{\partial }{\rho \partial \varphi }+\hat{z}\frac{\partial }{%
\partial z}$ and accordingly the Gradient, Divergence and Laplace operators
are in the standard form in 3-dimensional Euclidean cylindrical coordinates.
Also, $A_{\varphi }=A_{\varphi }(\rho ,z)$ is the $\varphi $ component of
the electromagnetic 4-potential. Finally, the $\gamma (\rho ,z)$ function
satisfies%
\begin{multline}
\partial _{\rho }\gamma =\frac{\rho \left[ \left( \partial _{\rho }\mathcal{E%
}+2\Phi ^{\star }\partial _{\rho }\Phi \right) \left( \partial _{\rho }%
\mathcal{E}^{\star }+2\Phi \partial _{\rho }\Phi ^{\star }\right) -\left(
\partial _{z}\mathcal{E}+2\Phi ^{\star }\partial _{z}\Phi \right) \left(
\partial _{z}\mathcal{E}^{\star }+2\Phi \partial _{z}\Phi ^{\star }\right) %
\right] }{4\left( \func{Re}\left( \mathcal{E}\right) +\left\vert \Phi
\right\vert ^{2}\right) ^{2}}-  \label{46} \\
\frac{\rho }{\func{Re}\left( \mathcal{E}\right) +\left\vert \Phi \right\vert
^{2}}\left[ \partial _{\rho }\Phi \partial _{\rho }\Phi ^{\star }-\partial
_{z}\Phi \partial _{z}\Phi ^{\star }\right] ,
\end{multline}%
and%
\begin{multline}
\partial _{z}\gamma =\frac{\rho \left[ \left( \partial _{\rho }\mathcal{E}%
+2\Phi ^{\star }\partial _{\rho }\Phi \right) \left( \partial _{z}\mathcal{E}%
^{\star }+2\Phi \partial _{z}\Phi ^{\star }\right) +\left( \partial _{z}%
\mathcal{E}+2\Phi ^{\star }\partial _{z}\Phi \right) \left( \partial _{\rho }%
\mathcal{E}^{\star }+2\Phi \partial _{\rho }\Phi ^{\star }\right) \right] }{%
4\left( \func{Re}\left( \mathcal{E}\right) +\left\vert \Phi \right\vert
^{2}\right) ^{2}}-  \label{47} \\
\frac{\rho }{\func{Re}\left( \mathcal{E}\right) +\left\vert \Phi \right\vert
^{2}}\left[ \partial _{\rho }\Phi \partial _{z}\Phi ^{\star }-\partial
_{z}\Phi \partial _{\rho }\Phi ^{\star }\right] .
\end{multline}%
The Harrison transformation acts on the Ernst potentials as 
\begin{equation}
\mathcal{E}^{\prime }=\frac{\mathcal{E}}{1-2\alpha ^{\star }\Phi -|\alpha
|^{2}\mathcal{E}},\qquad \Phi ^{\prime }=\frac{\alpha \mathcal{E}+\Phi }{%
1-2\alpha ^{\star }\Phi -|\alpha |^{2}\mathcal{E}},  \label{48}
\end{equation}%
where $\alpha $ is a complex constant.

As a seed solution, we consider the vacuum static axially symmetric CCS 
\begin{equation}
ds^{2}=-e^{-\frac{2M}{\sqrt{\rho ^{2}+z^{2}}}}dt^{2}+e^{\frac{2M}{\sqrt{\rho
^{2}+z^{2}}}}\left[ e^{-\frac{M^{2}\rho ^{2}}{(\rho ^{2}+z^{2})^{2}}}(d\rho
^{2}+dz^{2})+\rho ^{2}d\varphi ^{2}\right] ,  \label{49}
\end{equation}%
for which we extract 
\begin{equation}
\mathcal{E}=e^{-\frac{2M}{\sqrt{\rho ^{2}+z^{2}}}},\qquad \Phi =\chi =0.
\label{50}
\end{equation}%
Applying the Harrison transformation yields 
\begin{equation}
\mathcal{E}^{\prime }=\frac{1}{e^{\frac{2M}{\sqrt{\rho ^{2}+z^{2}}}}-|\alpha
|^{2}},\qquad \Phi ^{\prime }=\frac{\alpha }{e^{\frac{2M}{\sqrt{\rho
^{2}+z^{2}}}}-|\alpha |^{2}}.  \label{51}
\end{equation}%
Moreover, from $\mathcal{E}^{\prime }=f^{\prime }-|\Phi ^{\prime }|^{2}$, we
obtain 
\begin{equation}
f^{\prime }=\frac{e^{\frac{2M}{\sqrt{\rho ^{2}+z^{2}}}}}{\left( e^{\frac{2M}{%
\sqrt{\rho ^{2}+z^{2}}}}-|\alpha |^{2}\right) ^{2}},  \label{52}
\end{equation}%
and $\chi ^{\prime }=0$. Writing $\alpha =\alpha _{r}+i\alpha _{i}$, the
electromagnetic potentials are obtained to be 
\begin{equation}
A_{t}^{\prime }=\frac{\alpha _{r}}{e^{\frac{2M}{\sqrt{\rho ^{2}+z^{2}}}%
}-|\alpha |^{2}},\qquad \tilde{A}_{\varphi }^{\prime }=\frac{\alpha _{i}}{e^{%
\frac{2M}{\sqrt{\rho ^{2}+z^{2}}}}-|\alpha |^{2}}.  \label{53}
\end{equation}%
Since the transformed spacetime remains static ($\omega ^{\prime }=0$), Eq.~(%
\ref{44}) reduces to 
\begin{equation}
\nabla A_{\varphi }^{\prime }=\rho f^{\prime }{}^{-1}\hat{\varphi}\times
\nabla \tilde{A}_{\varphi }^{\prime },
\end{equation}%
such that integrating this expression yields 
\begin{equation}
A_{\varphi }^{\prime }=-\frac{2M\alpha _{i}z}{\sqrt{\rho ^{2}+z^{2}}},
\label{54}
\end{equation}%
up to a gauge constant. Thus, the electromagnetic potential one-form is
given by 
\begin{equation}
\mathbf{A}=\frac{\alpha _{r}}{e^{\frac{2M}{\sqrt{\rho ^{2}+z^{2}}}}-|\alpha
|^{2}}\,dt-\frac{2M\alpha _{i}z}{\sqrt{\rho ^{2}+z^{2}}}\,d\varphi .
\label{55}
\end{equation}%
Since the Harrison transformation leaves the metric function $\gamma $
invariant, i.e.\ $\gamma ^{\prime }=\gamma $, the charged static axially
symmetric spacetime generated by this transformation is given by 
\begin{equation}
ds^{\prime 2}=-\frac{e^{\frac{2M}{\sqrt{\rho ^{2}+z^{2}}}}}{\left( e^{\frac{%
2M}{\sqrt{\rho ^{2}+z^{2}}}}-|\alpha |^{2}\right) ^{2}}\,dt^{2}+\frac{\left(
e^{\frac{2M}{\sqrt{\rho ^{2}+z^{2}}}}-|\alpha |^{2}\right) ^{2}}{e^{\frac{2M%
}{\sqrt{\rho ^{2}+z^{2}}}}}\left\{ e^{-\frac{M^{2}\rho ^{2}}{(\rho
^{2}+z^{2})^{2}}}(d\rho ^{2}+dz^{2})+\rho ^{2}d\varphi ^{2}\right\} .
\label{56}
\end{equation}%
In spherical coordinates, the electromagnetic gauge-potential reads as 
\begin{equation}
\mathbf{A}=\frac{\alpha _{r}}{e^{\frac{2M}{r}}-|\alpha |^{2}}\,dt-2M\alpha
_{i}\cos \theta \,d\varphi ,  \label{57}
\end{equation}%
which yields the electromagnetic 2-form field 
\begin{equation}
\mathbf{F}=-\frac{2M\alpha _{r}}{r^{2}(1-|\alpha |^{2})^{2}\left( \cosh 
\frac{M}{r}+\frac{1+|\alpha |^{2}}{1-|\alpha |^{2}}\sinh \frac{M}{r}\right)
^{2}}\,dt\wedge dr+2M\alpha _{i}\sin \theta \,d\theta \wedge d\varphi .
\label{58}
\end{equation}%
Note that, the metric~(\ref{56}) can be rewritten as 
\begin{multline}
ds^{\prime 2}=-\frac{1}{(1-|\alpha |^{2})^{2}\left( \cosh \frac{M}{r}+\frac{%
1+|\alpha |^{2}}{1-|\alpha |^{2}}\sinh \frac{M}{r}\right) ^{2}}\,dt^{2}
\label{59} \\
+(1-|\alpha |^{2})^{2}\left( \cosh \frac{M}{r}+\frac{1+|\alpha |^{2}}{%
1-|\alpha |^{2}}\sinh \frac{M}{r}\right) ^{2}\left\{ e^{-\frac{M^{2}\sin
^{2}\theta }{r^{2}}}(dr^{2}+r^{2}d\theta ^{2})+r^{2}\sin ^{2}\theta
\,d\varphi ^{2}\right\} ,
\end{multline}%
such that upon introducing the radial rescaling 
\begin{equation*}
r=\frac{\tilde{r}}{|1-|\alpha |^{2}|},\qquad |\alpha |^{2}\neq 1,
\end{equation*}%
it takes the form 
\begin{multline}
ds^{\prime 2}=-\frac{1}{\left( \cosh \frac{\tilde{M}}{\tilde{r}}+\frac{%
1+|\alpha |^{2}}{1-|\alpha |^{2}}\sinh \frac{\tilde{M}}{\tilde{r}}\right)
^{2}}\,d\tilde{t}^{2}  \label{60} \\
+\left( \cosh \frac{\tilde{M}}{\tilde{r}}+\frac{1+|\alpha |^{2}}{1-|\alpha
|^{2}}\sinh \frac{\tilde{M}}{\tilde{r}}\right) ^{2}\left\{ e^{-\frac{\tilde{M%
}^{2}\sin ^{2}\theta }{\tilde{r}^{2}}}(d\tilde{r}^{2}+\tilde{r}^{2}d\theta
^{2})+\tilde{r}^{2}\sin ^{2}\theta \,d\varphi ^{2}\right\} ,
\end{multline}%
where we set $\tilde{t}=t/|1-|\alpha |^{2}|$ and $\tilde{M}=M|1-|\alpha
|^{2}|$. Also the electromagnetic field becomes 
\begin{equation}
\mathbf{F}=\frac{\tilde{Q}}{\tilde{r}^{2}\left( \cosh \frac{\tilde{M}}{%
\tilde{r}}+\frac{1+|\alpha |^{2}}{1-|\alpha |^{2}}\sinh \frac{\tilde{M}}{%
\tilde{r}}\right) ^{2}}\,d\tilde{t}\wedge d\tilde{r}+\tilde{P}\sin \theta
\,d\theta \wedge d\varphi ,  \label{61}
\end{equation}%
where the electric and magnetic charges are defined as 
\begin{equation*}
\tilde{Q}=-\frac{2\tilde{M}\alpha _{r}}{|1-|\alpha |^{2}|},\qquad \tilde{P}=%
\frac{2\tilde{M}\alpha _{i}}{|1-|\alpha |^{2}|},
\end{equation*}%
that satisfy 
\begin{equation}
\frac{\tilde{Q}^{2}+\tilde{P}^{2}}{4\tilde{M}^{2}}=\frac{|\alpha |^{2}}{%
(1-|\alpha |^{2})^{2}}.  \label{62}
\end{equation}%
Finally, introducing the parameter $\tilde{\mu}$ through 
\begin{equation}
\frac{1+|\alpha |^{2}}{1-|\alpha |^{2}}=\frac{\tilde{\mu}}{\tilde{M}},
\end{equation}%
and combining with Eq.~(\ref{62}), we obtain the relation 
\begin{equation}
\tilde{Q}^{2}+\tilde{P}^{2}=\tilde{\mu}^{2}-\tilde{M}^{2}.  \label{63}
\end{equation}%
Comparing these expressions with Eqs.~(\ref{9}), (\ref{25}), and~(\ref{26}),
we conclude that the solution obtained by directly solving the EM equations
coincides exactly with that generated via the Harrison transformation.

\section{Magnetic Harrison transformation}

The magnetic Harrison transformation introduces an external axisymmetric
magnetic field while preserving stationarity. For a purely magnetic
transformation characterized by the parameter $B$, the Ernst potentials
transform as 
\begin{equation}
\mathcal{E}^{\prime }=\frac{\mathcal{E}}{1-\frac{1}{4}B^{2}\mathcal{E}}%
,\qquad \Phi ^{\prime }=-\frac{\frac{1}{2}B\mathcal{E}}{1-\frac{1}{4}B^{2}%
\mathcal{E}}.  \label{T1}
\end{equation}%
The resulting spacetime represents a \emph{Melvin-type magnetized extension}
of the seed geometry. For the CCS seed metric, for which $\mathcal{E}=-\rho
^{2}e^{\frac{2M}{\sqrt{\rho ^{2}+z^{2}}}}$, one finds 
\begin{equation}
\mathcal{E}^{\prime }=-\frac{\rho ^{2}e^{\frac{2M}{\sqrt{\rho ^{2}+z^{2}}}}}{%
\Lambda (\rho ,z)},\qquad \Phi ^{\prime }=\frac{\frac{1}{2}B\rho ^{2}e^{%
\frac{2M}{\sqrt{\rho ^{2}+z^{2}}}}}{\Lambda (\rho ,z)},  \label{T2}
\end{equation}%
where 
\begin{equation}
\Lambda (\rho ,z)=1+\frac{1}{4}B^{2}\rho ^{2}e^{\frac{2M}{\sqrt{\rho
^{2}+z^{2}}}}.  \label{T4}
\end{equation}%
The corresponding magnetized spacetime metric takes the form 
\begin{equation}
ds^{\prime 2}=-\Lambda ^{2}e^{-\frac{2M}{\sqrt{\rho ^{2}+z^{2}}}%
}\,dt^{2}+\Lambda ^{2}e^{\frac{2M}{\sqrt{\rho ^{2}+z^{2}}}}\left\{ e^{-\frac{%
M^{2}\rho ^{2}}{(\rho ^{2}+z^{2})^{2}}}(d\rho ^{2}+dz^{2})+\frac{\rho ^{2}}{%
\Lambda ^{4}}d\phi ^{2}\right\} ,  \label{T3}
\end{equation}%
and the electromagnetic four-potential generated by the magnetic Harrison
transformation is given by 
\begin{equation}
\mathbf{A}=\frac{B\rho ^{2}e^{\frac{2M}{\sqrt{\rho ^{2}+z^{2}}}}}{2\Lambda
(\rho ,z)}\,d\phi .  \label{T5}
\end{equation}%
In spherical coordinates, the line element can be written as 
\begin{equation}
ds^{\prime 2}=-e^{-\frac{2M}{r}}\left( 1+\frac{B^{2}r^{2}\sin ^{2}\theta }{4}%
e^{\frac{2M}{r}}\right) ^{2}\,dt^{2}+\left( 1+\frac{B^{2}r^{2}\sin
^{2}\theta }{4}e^{\frac{2M}{r}}\right) ^{2}e^{\frac{2M}{r}}\left\{ e^{-\frac{%
M^{2}\sin ^{2}\theta }{r^{2}}}\left( dr^{2}+r^{2}d\theta ^{2}\right) +\frac{%
r^{2}\sin ^{2}\theta \,d\phi ^{2}}{\left( 1+\frac{B^{2}r^{2}\sin ^{2}\theta 
}{4}e^{\frac{2M}{r}}\right) ^{4}}\right\} ,
\end{equation}%
such that in the limit $M\rightarrow 0$, the metric smoothly reduces to the
Melvin magnetic universe, while in the limit $B\rightarrow 0$ the original
vacuum CCS solution is recovered.

\section{Ehlers transformation}

Following the Harrison transformation, we now apply the so-called Ehlers
transformation \cite{Ehlers1957}, which in terms of the Ernst potentials is
defined as 
\begin{equation}
\mathcal{E}^{\prime }=\frac{\mathcal{E}}{1+ic\mathcal{E}},\qquad \Phi
^{\prime }=\frac{\Phi }{1+ic\mathcal{E}},  \label{64}
\end{equation}%
where $c$ is a real parameter. The seed solution is given by Eq.~(\ref{49}),
and the corresponding Ernst potentials are provided in Eqs.~(\ref{41}) and (%
\ref{43}). Since the seed spacetime is vacuum, $\Phi =0$, and Eq.~(\ref{64})
immediately yields 
\begin{equation}
\mathcal{E}^{\prime }=\frac{f}{1+icf},  \label{65}
\end{equation}%
together with 
\begin{equation}
\Phi ^{\prime }=0.  \label{66}
\end{equation}%
Equation~(\ref{65}) can be rewritten as 
\begin{equation}
\mathcal{E}^{\prime }=\frac{f(1-icf)}{1+c^{2}f^{2}},  \label{67}
\end{equation}%
from which one identifies 
\begin{equation}
\chi ^{\prime }:=\Im \!\left( \mathcal{E}^{\prime }\right) =-\frac{cf^{2}}{%
1+c^{2}f^{2}},  \label{68}
\end{equation}%
and 
\begin{equation}
f^{\prime }:=|\Phi ^{\prime }|^{2}+\Re \!\left( \mathcal{E}^{\prime }\right)
=\frac{f}{1+c^{2}f^{2}}.  \label{69}
\end{equation}%
Furthermore, using Eq.~(\ref{45}), the twist potential satisfies 
\begin{equation}
\nabla \omega ^{\prime }:=-\frac{\rho }{f^{\prime 2}}\hat{\varphi}\times %
\left[ \nabla \chi ^{\prime }+2\,\Im \!\left( \Phi ^{\prime \star }\nabla
\Phi ^{\prime }\right) \right] =\frac{\rho (1+c^{2}f^{2})^{2}}{f^{2}}\hat{%
\varphi}\times \nabla \!\left( \frac{cf^{2}}{1+c^{2}f^{2}}\right) ,
\label{70}
\end{equation}%
which can be integrated to give 
\begin{equation}
\omega ^{\prime }=-\frac{4cMz}{\sqrt{\rho ^{2}+z^{2}}}.  \label{71}
\end{equation}%
The remaining field equations, Eqs.~(\ref{46}) and (\ref{47}), are satisfied
with the same metric function $\gamma (\rho ,z)$ as the seed solution.
Consequently, the transformed line element takes the form 
\begin{multline}
ds^{\prime 2}=-\frac{1}{e^{\frac{2M}{\sqrt{\rho ^{2}+z^{2}}}}+c^{2}e^{-\frac{%
2M}{\sqrt{\rho ^{2}+z^{2}}}}}\left( dt+\frac{4cMz}{\sqrt{\rho ^{2}+z^{2}}}%
\,d\varphi \right) ^{2}  \label{72} \\
+\left( e^{\frac{2M}{\sqrt{\rho ^{2}+z^{2}}}}+c^{2}e^{-\frac{2M}{\sqrt{\rho
^{2}+z^{2}}}}\right) \left[ e^{-\frac{M^{2}\rho ^{2}}{(\rho ^{2}+z^{2})^{2}}%
}(d\rho ^{2}+dz^{2})+\rho ^{2}d\varphi ^{2}\right] .
\end{multline}%
In spherical coordinates, the metric becomes 
\begin{equation}
ds^{\prime 2}=-\frac{1}{e^{\frac{2M}{r}}+c^{2}e^{-\frac{2M}{r}}}\left(
dt+4cM\cos \theta \,d\varphi \right) ^{2}+\left( e^{\frac{2M}{r}}+c^{2}e^{-%
\frac{2M}{r}}\right) \left[ e^{-\frac{M^{2}\sin ^{2}\theta }{r^{2}}%
}(dr^{2}+r^{2}d\theta ^{2})+r^{2}\sin ^{2}\theta \,d\varphi ^{2}\right] ,
\label{73}
\end{equation}%
in which after introducing $r=\tilde{r}/(1+c^{2})$, $t=(1+c^{2})\tilde{t}$, $%
M=\tilde{M}(1+c^{2})$, and $\ell =2c\tilde{M}$, the metric can be cast into
the form 
\begin{multline}
ds^{\prime 2}=-\left[ \cosh \!\left( \frac{2\tilde{M}}{\tilde{r}}\right) +%
\frac{1-c^{2}}{1+c^{2}}\sinh \!\left( \frac{2\tilde{M}}{\tilde{r}}\right) %
\right] ^{-1}\left( d\tilde{t}+2\ell \cos \theta \,d\varphi \right) ^{2}
\label{74} \\
+\left[ \cosh \!\left( \frac{2\tilde{M}}{\tilde{r}}\right) +\frac{1-c^{2}}{%
1+c^{2}}\sinh \!\left( \frac{2\tilde{M}}{\tilde{r}}\right) \right] \left[
e^{-\frac{\tilde{M}^{2}\sin ^{2}\theta }{\tilde{r}^{2}}}(d\tilde{r}^{2}+%
\tilde{r}^{2}d\theta ^{2})+\tilde{r}^{2}\sin ^{2}\theta \,d\varphi ^{2}%
\right] .
\end{multline}%
This spacetime is a vacuum solution of the Einstein equations in the Weyl
class and admits two commuting Killing vectors, $\xi _{(t)}=\partial _{t}$
and $\xi _{(\phi )}=\partial _{\phi }$. Although $g_{t\phi }\neq 0$, the
spacetime is not rotating. Indeed, the Komar angular momentum, 
\begin{equation}
J_{\mathrm{Komar}}:=\frac{1}{16\pi }\int_{S_{\infty }^{2}}\nabla ^{\mu }\xi
_{(\phi )}^{\nu }\,dS_{\mu \nu },  \label{75}
\end{equation}%
vanishes asymptotically. This can be verified by using 
\begin{equation}
g_{t\phi }=-2\ell \left[ \cosh \!\left( \frac{2\tilde{M}}{\tilde{r}}\right) +%
\frac{1-c^{2}}{1+c^{2}}\sinh \!\left( \frac{2\tilde{M}}{\tilde{r}}\right) %
\right] ^{-1}\cos \theta \sim -2\ell \cos \theta ,\qquad \tilde{r}%
\rightarrow \infty ,  \label{76}
\end{equation}%
that yields (In the asymptotic region only the ($t,\tilde{r}$) components
contribute) 
\begin{equation}
J_{\mathrm{Komar}}\sim \int_{0}^{\pi }\cos \theta \sin \theta \,d\theta =0.
\label{77}
\end{equation}%
Furthermore, the one-form associated with the timelike Killing vector, 
\begin{equation}
\xi _{(t)}=g_{t\mu }dx^{\mu }=-F(r)^{-1}\left( dt+2\ell \cos \theta \,d\phi
\right) ,  \label{78}
\end{equation}%
is not hypersurface orthogonal, and its exterior derivative yields a
nonvanishing twist, 
\begin{equation}
\omega =\star (\xi _{(t)}\wedge d\xi _{(t)})\neq 0.  \label{79}
\end{equation}%
This nonzero twist signals the presence of a \emph{gravitomagnetic monopole}
rather than ordinary rotation. The corresponding NUT charge can be defined
invariantly as the Komar dual mass, 
\begin{equation}
N=\frac{1}{8\pi }\int_{S_{\infty }^{2}}\star d\xi _{(t)},  \label{80}
\end{equation}%
which, upon evaluation using Eq.~(\ref{76}), yields 
\begin{equation}
N=\ell .  \label{81}
\end{equation}%
Therefore, the parameter $\ell $ is identified invariantly as the NUT
charge, while the Komar angular momentum vanishes. The Ehlers transformation
thus generates a \emph{swirling (NUT-type)} extension rather than a rotating
one.

\subsection{Magnetized swirling (NUT-type) spacetime}

In this section, we apply the magnetic Harrison transformation to the vacuum
swirling (NUT-type) spacetime given in Eq.~(\ref{74}). The resulting
geometry describes a \emph{magnetized swirling (NUT-type) spacetime}, whose
line element takes the form 
\begin{multline}
ds^{\prime 2}=-\frac{\Lambda ^{2}}{\left[ \cosh \!\left( \frac{2M}{r}\right)
+\frac{1-c^{2}}{1+c^{2}}\sinh \!\left( \frac{2M}{r}\right) \right] }\left(
dt+2\ell \cos \theta \,d\varphi \right) ^{2}  \label{82} \\
+\left[ \cosh \!\left( \frac{2M}{r}\right) +\frac{1-c^{2}}{1+c^{2}}\sinh
\!\left( \frac{2M}{r}\right) \right] \Lambda ^{2}\left[ e^{-\frac{M^{2}\sin
^{2}\theta }{r^{2}}}(dr^{2}+r^{2}d\theta ^{2})+\frac{r^{2}\sin ^{2}\theta
\,d\varphi ^{2}}{\Lambda ^{4}}\right] .
\end{multline}%
The electromagnetic four-potential generated by the Harrison transformation
is given by Eq.~(\ref{T5}), where the function $\Lambda (\rho ,z)$ is
defined in Eq.~(\ref{T4}). Although the Harrison transformation introduces a
magnetic field, it does not alter the Komar dual mass, hence the NUT charge
remains $\ell $.

\section{Conclusion}

In this work, we have presented a comprehensive construction of charged,
magnetized, and swirling extensions of the axially symmetric exponential
metric, a vacuum Weyl solution closely related to the CCS. Starting from the
EM action, we obtained an exact charged axially symmetric exponential metric
by solving the coupled field equations for a dyonic monopole source. The
resulting geometry interpolates between the vacuum CCS and the extremal RN
spacetime and exhibits asymptotic behavior consistent with a deformed
isotropic RN solution. Using the Ernst formalism, we demonstrated that the
same charged solution can be generated via the Harrison transformation,
thereby confirming the consistency and robustness of the solution generating
approach. We further applied the magnetic Harrison transformation to
construct Melvin-type magnetized versions of both the vacuum and charged
spacetimes, recovering the Melvin magnetic universe in the appropriate
limit. Employing the Ehlers transformation, we generated a stationary but
nonrotating swirling geometry characterized by a nonvanishing NUT charge.
Despite the presence of an off-diagonal $g_{t\phi }$ component, the Komar
angular momentum vanishes, while the Komar dual mass identifies the NUT
parameter invariantly. This confirms that the transformation induces a
gravitomagnetic monopole rather than ordinary rotation. A subsequent
magnetic Harrison transformation yielded a magnetized swirling (NUT-type)
spacetime, combining electric, magnetic, and gravitomagnetic features in a
single exact solution. Altogether, our results illustrate how the axially
symmetric exponential metric serves as a versatile seed for generating a
rich family of exact EM solutions. These spacetimes provide a unified
framework for studying the interplay between anisotropy, electromagnetic
charges, external magnetic fields, and gravitomagnetic effects, and may
offer useful testbeds for further investigations of geodesic motion,
lensing, and thermodynamic properties in non-spherically symmetric
gravitational fields.

\end{document}